\documentclass{aa}
\usepackage{graphicx}

\begin{document}

\title{On the Statistical Significance of the Bulk Flow Measured by the PLANCK Satellite}

\author{F.~Atrio-Barandela}
\institute{F{\'\i}sica Te\'orica, Universidad de Salamanca,
37008 Salamanca, Spain; atrio@usal.es}

\date{}

\abstract{
A recent analysis of data collected by the Planck satellite detected a net  dipole at 
the location of X-ray selected galaxy clusters, corresponding to a large-scale bulk flow 
extending at least to $z\sim 0.18$, the median redshift of the cluster sample. The amplitude 
of this flow, as measured with Planck, is consistent with earlier findings based on data from 
the Wilkinson Microwave Anisotropy Probe (WMAP). However, the uncertainty assigned to 
the dipole by the Planck team is much larger than that found in the WMAP studies, leading 
the authors of the Planck study to conclude that the observed bulk flow is not statistically
significant. \\
We here show that two of the three implementations of random sampling used in the error 
analysis of the Planck study lead to systematic overestimates in the uncertainty of the 
measured dipole. 
Random simulations of the sky do not take into account that the actual realization of the sky 
leads to filtered data  that have a 12\% lower root-mean-square dispersion than the average 
simulation.  Using rotations around the Galactic pole (the Z axis), increases 
the uncertainty of the X and Y components of the dipole and artificially reduces the 
significance of the dipole detection from 98-99\% to less than 90\% confidence. 
When either effect is taken into account, the corrected errors agree with those 
obtained using random distributions of clusters on Planck data, 
and the resulting statistical significance of the 
dipole measured by Planck is consistent with that of the WMAP results. 
}



\maketitle

\section{Introduction.\label{sec:intro}}

The intrinsic difficulty of determining peculiar velocities from galaxy redshifts and 
distance indicators led Kashlinsky \& Atrio-Barandela (2000) to propose an alternative 
method of probing the velocity field on large scales.  Galaxy clusters leave an imprint 
on the Cosmic Microwave Background (CMB) in the form of distortions in the CMB 
black-body spectrum caused by the Sunyaev-Zel`dovich (SZ) effect 
(Sunyaev \& Zel`dovich 1970, 1972). Two different mechanisms contribute to the SZ 
effect: the thermal component (TSZ) is caused by the thermal motions of electrons in 
the potential wells of clusters, whereas the kinematic component (KSZ) is due to the 
motion of the cluster as a whole. We noted that any bulk flow of clusters would produce 
a dipole in the anisotropy temperature in the direction of clusters. Since this signal is 
small compared to the sampling variance of the intrinsic CMB dipole at the same positions, 
we proposed to use the statistical properties of the CMB data to filter out the dominant 
cosmological component, thereby increasing the signal-to-noise ratio of any contribution 
from a bulk-flow dipole.  

In Atrio-Barandela et al.\ (2008) and Kashlinsky et al.\ (2008, 2009) we presented results 
of our application of this method to Wilkinson Microwave Anisotropy Probe (WMAP) 3-year 
data, later extended to WMAP 5-year  and 7-year data (Kashlinsky et al.\ 2010; Kashlinsky, 
Atrio-Barandela \& Ebeling 2011). For a sample of $\sim$700 X-ray-selected clusters, we 
detected a persistent dipole, measured at the cluster positions within apertures 
(of 25\arcmin\ radius) that contain zero TSZ monopole. The dipole is roughly aligned with the 
CMB dipole and can be traced to cluster redshifts exceeding $z\sim 0.2$; its amplitude 
correlates with that of the monopole within apertures of 10\arcmin\ radius.  We interpreted 
this signal, associated exclusively with clusters, as evidence of a large-scale flow of 
amplitude $\simeq$ 600--1000 km s$^{-1}$ that could encompass the whole observable 
horizon. A ``Dark Flow" of this 
amplitude, if real, would be equivalent to the all-sky  CMB dipole being 
primarily of primordial origin, intrinsic to last scattering surface.  

Our theoretical and numerical estimates indicated that our dipole detection  is significant 
at the 99.4\% confidence level. However, independent confirmation of this result is still 
lacking, and several studies have challenged our results. Keisler (2009) confirmed the 
existence of the dipole detected by us, but claimed that it was not statistically significant. 
It was shown though (Atrio-Barandela et al.\ 2010) that Keisler neglected to subtract the 
dipole outside the Galactic mask, causing the error bars of his measurement to be 
overestimated. 
More recently, Osborne et al.\ (2011) and Mody \& Hajian (2012) did not find bulk 
flows in WMAP data using filtering schemes different from ours. However, 
both teams of authors implicitly assumed that clusters have the same angular extent in the 
original data as in the filtered maps, and that the electron pressure and 
the electron density in clusters follow the same radial profile. Both assumptions 
are incorrect and render these 
filters insensitive to bulk flows. In Atrio-Barandela et al.\ (2013) we demonstrated that 
correct implementation of either of these alternative filtering schemes leads to results 
that are consistent with ours. Two recent studies using galaxies, rather than clusters, 
to search for  the KSZ signature of a large-scale bulk flow in CMB data have also challenged 
our findings. The CMB-galaxy cross-correlation study by Li, Zhang \& Chen (2012) found 
a statistically anisotropic component in the CMB on scales out to $z\sim 0.2$, but ruled 
out the Dark Flow  as its cause, due to the very high galaxy velocities implied by this 
interpretation. We note though that peculiar velocities in perfect agreement with our 
findings would result if their cross-correlation signal were dominated by cluster galaxies 
rather than field galaxies. Lavaux, Afshordi \& Hudson  (2013)  aimed to use the 
KSZ effect from the plasma halo of galaxies to probe the peculiar velocity field and found 
results on small scales (${<}50h^{-1}$ Mpc) that are consistent with those obtained using 
galaxy distance indicators. On scales of ${\sim}500h^{-1}$ Mpc, however, their analysis 
leads to an upper limit of 470 km s$^{-1}$ (95\% significance) for the bulk-flow velocity, 
in conflict with the Dark Flow amplitude measured by us. It should be noted though that 
these conclusions depend sensitively on the spatial distribution of ionized gas in and 
around galaxies, which is presently ill-constrained by observations.

A conclusive determination of the Dark Flow is expected to emerge from an 
authoritative study based on CMB data collected by the Planck satellite. 
With its higher resolution, better frequency coverage and lower noise levels Planck is 
much better suited for  studies of the SZ effect than WMAP. A comparison between the 
dipole measurements from the two missions will also allow a much-needed consistency 
check, since differences between the instruments and scanning strategies result in data 
with different systematics. In this context, the Planck Collaboration has recently 
reported (Planck Collaboration. Planck 2013 results. XXVII. 2013)
the aberration and modulation effects due to the motion of the Sun with respect
to the CMB frame and found a significant agreement with the Doppler boosting
of the CMB monopole. However, this measurement does not constrain the motion 
of the Local Group but the amplitude of its component in the direction of 
the solar motion.  They have also reported the
first Planck-based KSZ measurement of the peculiar motions of clusters (Planck 
Collaboration 2013, Planck Intermediate Results XIII, hereafter PIR-13) using their 
own catalog of X-ray selected clusters (Piffaretti et al.\ 2011) and concluded
that the Planck constraints on peculiar velocities were consistent with
the concordance $\Lambda$CDM model.  

In this paper we analyze and evaluate the error estimators employed by PIR-13. We begin 
by summarizing the PIR-13 results and methods in Sec.~2. Sec.~3 briefly reviews the 
formalism regarding error estimation developed by Atrio-Barandela et al.\ (2010), before 
we discuss in detail  two problematic implementations of error estimation used  in PIR-13, 
rotations and numerical simulations, in Sec.~4 and 5, respectively. 
We present our conclusions in Sec.~6.

\section{Results and methods of the first Planck bulk-flow study}
\label{sec:planck}

The PIR-13 results characterize the average peculiar velocities of clusters at different 
depths using three complementary diagnostics: the radial peculiar-velocity average, 
the root-mean-square (rms) velocity, and the bulk-flow velocity. The underlying analysis 
employs three different filtering schemes, including the one introduced by us 
(Kashlinsky et al.\ 2009). The study's findings for all three diagnostics mentioned 
above are consistent with the very low $\Lambda$CDM predictions for the average 
peculiar velocities of galaxy clusters. Specifically, the PIR-13 abstract asserts 
that ``there is no detection of [a] bulk flow as measured in any comoving 
sphere extending to the maximum redshift covered by the cluster sample''.  
To fully appreciate the nature and significance of this assertion, it is 
critical to understand that, when using our filtering scheme, PIR-13  
detects the same dipole signal at cluster positions as previously found by us in 
WMAP data (PIR-13, Fig 10). 
\textit{What renders their bulk-flow detection insignificant is thus not 
the absence of a dipole signal, but the uncertainty assigned to it.} 
Before analyzing PIR-13's error estimates in more detail in the following section, 
we here briefly discuss several key aspects of their work.

The PIR-13 study is based on the Meta Catalogue of X-ray detected Clusters of galaxies 
(MCXC; Piffaretti et al.\ 2011), a compilation of  publicly available X-ray cluster samples 
drawn primarily from ROSAT All-Sky Survey projects, but also including mostly low-mass 
systems from serendipitous X-ray cluster surveys. Comprising 1750 clusters and  rich galaxy 
groups, the MCXC is comparable in size to our proprietary catalog of 1558 X-ray selected 
clusters, but contains more low-luminosity systems and fewer very X-ray luminous clusters. 
By excluding MCXC entries close to bright point sources in any single-frequency Planck map, 
clusters lying in regions of high Galactic emission, and systems with estimated masses below 
$10^{13}M_\odot$, the sample used by PIR-13 is reduced to 1405 clusters. PIR-13 conducted 
their analysis using a CMB map that is the weighted average of the single-frequency Planck 
maps. This so-called 2D Internal Linear Combination (2D-ILC) map  introduces additional 
masking, resulting in a final sample of 1321 MCXC clusters. For comparison, our sample 
was limited to the 1205 clusters with X-ray luminosities 
$L_X[{\rm 0.1-2.4\,\,keV}]\ge 0.5\times 10^{44}$ erg s$^{-1}$, and our results 
were obtained from 796 clusters that fall outside the extended KQ75 mask.

Comparing the WMAP and Planck cluster dipoles allows us to identify and account for 
previously undiagnosed systematic effects.  First, differences in scanning strategy,
angular resolution, foreground residuals and other systematics could give
different error bars in WMAP and Planck even if the measured dipole values are 
the same. Second, a larger cluster catalog must improve the significance of 
the result if the cluster mass range is comparable. Third, the TSZ contribution 
can be determined and monitored much better with Planck data by comparing measurements 
at frequencies above and below the TSZ null at 217 GHz. Then, any correlation between 
the TSZ monopole within $10\arcmin$ (radius) apertures and the KSZ dipole within 
$25\arcmin$ apertures (Kashlinsky et al.\ 2010, Kashlinsky et al.\ 2011) can also 
be established better with Planck than with WMAP. Nevertheless, a direct 
comparison of the results could present some difficulties. In our WMAP-based 
analysis, the dipole signal was measured within discs of radius $25\arcmin$ around 
each cluster, corresponding to the zero monopole aperture and, due to its higher 
resolution, the zero monopole aperture in Planck and in WMAP may be different. 
In addition, the correlation between the SZ monopole and 
dipole observed by us indicates that low-mass systems contribute little to the 
signal. A larger fraction of low-mass clusters in the MCXC could thus dilute the 
dipole, although we note in this paper the additional reasons for which PIR-13 failed 
to detect a significant cluster dipole, as we shall elaborate below.

PIR-13  applied several filtering schemes in Legendre space. While the actual filters  
differ from those previously used and described in the literature, the PIR-13 filter 
implementations share the drawbacks of the filters used by Osborne et al.\ (2011) 
and Mody \& Hajian (2012). Specifically, the PIR-13 study assumes that clusters have 
the same size in the real as in the filtered data, and adopts isothermal cluster models 
to calibrate these filters. Both of these assumptions, however, render filters insensitive 
to bulk flows, as shown by Atrio-Barandela et al.\ (2013).  In addition, PIR-13 uses the 
Aperture Photometry filter, first introduced  in the context of bulk-flow measurements 
by Kashlinsky \& Atrio-Barandela (2000). Contrary to other filters, the Aperture Photometry 
filter operates in real space: the intrinsic CMB signal is removed by subtracting the 
temperature averaged within an annulus (defined by radii $\theta_{in},\theta_{out}$) 
from the temperature averaged within the inner disc of radius $\theta_{in}$. While this 
process indeed removes the CMB, it also removes part of the SZ anisotropy of the 
cluster. In order to estimate the missing fraction, the profiles of the electron pressure 
and electron density need to be known for each cluster. Assuming clusters to be
isothermal simplifies 
the task, but biases the result. If the electron density falls less rapidly than the electron 
pressure, then the KSZ component is removed more efficiently than the TSZ component, 
reducing the cluster dipole. In addition, all cluster annuli must have the same radii if the 
CMB is to be removed uniformly across the sky, but PIR-13 defines different radii for 
each cluster, causing the residual CMB to vary from cluster to cluster. A detailed 
discussion of the resulting complications will be given in a forthcoming paper.

Finally, and most importantly, the PIR-13 analysis uses three methods to estimate the 
uncertainties of their measurements: (1) dipole measurements performed on the filtered 
maps at randomized cluster positions, (2) dipole measurements performed on random 
realizations of the filtered data, but at the real cluster positions, and (3) 
dipole measurements 
performed by rotating the template of cluster positions on the real data. PIR-13 offers no 
discussion of these estimators' relative efficiency or biases, which is surprising in view of 
the fact  that the resulting error estimates differ greatly.  When using (1), the significance 
of the KSZ dipole found by PIR-13 using Planck data is higher than the one found by us 
from WMAP data, but it is lower when using methods (2) and (3). A bias inherent in method 
(2) is well established and due to the power spectrum of the residual CMB component in 
the actual realization of the sky being about 12\% smaller than that of the average random 
realization of the sky (Atrio-Barandela et al.\ 2010). In the following sections, 
we discuss key aspects of the error computation as well as the 
discrepancies of the error estimation methods (2) and (3) with method (1) in more detail.

\section{On the Error Bars of the Bulk Flow Measurement}
\label{sec:error_theory}

Since clusters subtend a small solid angle on the sky, sampling variance is the largest 
source of uncertainty in the determination of bulk flows (Kashlinsky \& Atrio-Barandela 2000).
Random dipoles from the cosmological CMB component have much larger amplitude than 
any KSZ dipole caused by the bulk motion of a cluster sample. To suppress this noise term, 
we suggested to filter out the CMB signal using the statistical properties of the temperature 
anisotropy field. In addition, contributions from the Galaxy and point sources are removed 
by application of suitably defined masks. We assume use of the extended WMAP 7yr mask 
(the KQ75 mask) in the following description of the convolution process that creates the 
filtered maps: 
\begin{enumerate}

\item{} Each CMB map is multiplied by the KQ75 mask. Monopole, dipole, and quadrupole 
are removed from the remaining pixels.

\item{} The resulting CMB map is expanded into spherical harmonics with coefficients 
$(a^{sky}_{\ell m})$. The radiation power spectrum is 
$C_\ell^{sky}=\sum_m |a^{\rm sky}_{\ell m}|^2/f_{\rm sky}$, where the division by 
$f_{\rm sky}$, the fraction of the sky outside the KQ75 mask, accounts for the power 
removed by the masking process.

\item{} The power spectrum is filtered, and a Legendre transformation back to 
real space is applied: $\Delta T^{\rm fil}=\sum a^{sky}_{\ell m}F_\ell Y_{\ell m}$. 
We used the following Wiener-type filter: 
$F_\ell =[C_\ell^{\rm sky}-C_\ell^{\rm \Lambda CMB}B_\ell^2]/C_\ell^{\rm sky}$, 
where $B_\ell$ is the antenna beam, and $C_\ell^{\rm \Lambda CMB}$ is the 
theoretical CMB radiation power spectrum of the concordance model that best fits the data.

\item{} Finally, the filtered map $\Delta T^{\rm fil}$ is multiplied by the mask and 
any monopole and dipole introduced by the filtering process are removed from the 
remaining pixels. 
\end{enumerate}

In Atrio-Barandela et al (2010) we presented a comprehensive discussion of 
the errors associated with our method and showed that the variance of the monopole 
and dipole components is 
\begin{equation}
{\rm Var}(a_0)=\frac{\sigma^2_{\rm CMB,fil}}{N_{\rm cl}}+
\frac{\sigma^2_{\rm noise,fil}}{N_{\rm pix}}\qquad
{\rm Var}(a_{1i})=\frac{{\rm Var}(a_0)}{\langle n_i^2\rangle}
\label{eq:var}
\end{equation}
with $(n_i)=(X,Y,Z)=(\cos l\cos b,\sin l\cos b,\sin b)$ being the direction cosines 
of the cluster positions on the sky. In this expression, $\sigma_{\rm noise,fil}$ and 
$\sigma_{\rm CMB,fil}$ are the residual CMB and noise in the filtered maps. The 
noise is uncorrelated from pixel to pixel, whereas the correlation function of the 
residual CMB noise crosses zero at $\theta\sim 1^0$. Therefore, the sample 
variance of the contribution from the residual CMB scales approximately as the 
number of clusters $N_{\rm cl}$, while the  noise contribution scales with the number 
of pixels $N_{\rm pix}$. 

When estimating measurement uncertainties, we have to take into account  (a) that 
clusters are not randomly distributed in the sky and that any anisotropy in their distribution  
could increase the errors, and (b) that, in addition to instrumental noise, filtered maps also 
contain unknown residuals of foreground emission and other systematics that cannot be 
precisely modeled.  To take into account all sources of statistical uncertainty, error bars 
would have to be estimated using both the filtered data and the actual 
template of clusters in the sky. This is not possible (see also Sec.~\ref{sec:error_rot}). 
However, errors computed using random templates on the filtered data showed a behavior 
very close to the prediction of eq.~(\ref{eq:var}), indicating that foreground residuals 
and other possible systematics can be safely neglected (Atrio-Barandela et al.\ 2010; 
Kashlinsky, Atrio-Barandela \& Ebeling 2012). 
To further explore item (a), the impact of anisotropies in the cluster distribution, 
we will here consider two templates: T1, defined as the 796 clusters in our catalog with  
$L_{\rm X}\ge 0.5\times 10^{44}$ erg s$^{-1}$ in the ROSAT broad band (0.1--2.4 keV), 
and T2, a subset of T1, comprising only the 327 most X-ray luminous clusters 
($L_{\rm X}\ge 2.0\times 10^{44}$ erg s$^{-1}$) at $z\le 0.3$. 

We compute the dipole by assigning to 
all clusters the same angular radius of 25\arcmin, which roughly corresponds 
to the aperture within which the monopole vanishes in the filtered maps
constructed with WMAP 7-year data.  At {\sc Healpix} resolution $N_{\rm side}=512$ 
(Gorski et al.\ 2005), samples T1 and T2 occupy 32,700 and 13,500 pixels, respectively 
(about 1\% and 0.4\% of the sky). Since gravitational instability drives low-mass systems 
towards the more massive clusters, we expect the distribution of the T1 sample to be less 
isotropic than that of the T2 template, offering the possibility to quantify the impact of 
anisotropies on the measurement errors. Using those templates and WMAP 7-year data 
from the W1 Differencing Assembly (DA), we find  $\sigma_{\rm CMB,fil}\approx 30\mu$K 
and $\sigma_{\rm noise,fil}\approx 75\mu$K. Given the relatively modest number of 
clusters in our samples T1 and T2, eq.~(\ref{eq:var}) then indicates that the residual 
CMB in the filtered map dominates the errors, as suggested independently by 
Keisler et al.\ (2009) and Kashlinsky et al.\ (2010). Simulations of filtered maps thus 
need to contain only the CMB, and can neglect noise, foreground residuals, and other 
systematics. 

\section{Error estimates from rotated cluster templates}
\label{sec:error_rot}

As mentioned before, a  rigorous computation of errors would require measuring  
dipoles at random locations in the filtered data, outside the cluster apertures, but 
using the same template that describes the real cluster positions used in the analysis. 
Because of the complex geometry of the mask, this is not possible. 
In Atrio-Barandela et al.\ (2010) 
we discussed the methods labelled (1) and (2) in Sec.~\ref{sec:planck}: option (1) entails 
measuring dipoles in the filtered data, using random cluster templates that have no overlap 
with the real one, whereas option (2)  measures dipoles in simulated data, but at the position 
of the real clusters. When applying method (1), we account for the contributions to the 
measurement error from foreground residuals and other systematics as far as they are 
known; with method (2), we account for the effect of anisotropic distribution of our cluster 
template on the sky as detailed later in this section. In our simulations we do not find any 
systematic differences between results obtained for real and simulated cluster templates 
beyond cosmic variance. This is understandable since clusters, at random or real positions, 
are, by design, selected outside the Galactic plane, and hence all templates share the 
large-scale inhomogeneity of the mask. 

In PIR-13 a third method was introduced: (3) dipoles were measured in the filtered 
data at locations obtained {\it by rotating the cluster template around the Z axis 
in steps of one degree.} 
The Planck Collaboration expected this method to be optimal, since it includes all possible 
systematics of both the filtered data and the cluster template, including anisotropies and 
angular correlations. Unfortunately, this method is very inefficient and yields systematically 
larger errors than homogeneous random sampling. For one, the fixed step size of one 
degree allows only 359 different measurements and thus yields very few independent 
dipole estimates. In addition, a rotation does not move the clusters within the template alike. 
For instance, under rotation about the Z axis, the Coma cluster, located at $b=88^\circ$, 
never moves by more than four degrees from its initial position.  Finally, for small rotation 
angles, nearby clusters that are very extended on the sky will continue to contribute to the 
random dipole, correlating the measurements. As a result, the space of all possible dipoles 
is poorly sampled; the measured random dipoles overpopulate the tail of the distribution, 
leading to overestimated uncertainties that dilute the significance of any real result. 

In the following we discuss these systematic effects in more detail.

\subsection{Artificial correlations}

Fig.~\ref{fig:rotations} shows the monopole, dipole, and angular direction of the 
dipole measured by rotating the T1 template (796 clusters) in the same manner 
as prescribed by PIR-13. We show results obtained with the  W4 DA since it yields 
the largest dipole measured in the WMAP W-band, making it easier to identify 
any trends and biases. While, by design, all clusters in the unrotated T1 template 
fall outside the KQ75 mask, this is no longer the case after rotation\footnote{In principle 
the cluster template could of course be rotated about any axis but, due to the symmetry 
of the KQ75 mask with respect to the plane of the Galaxy, the choice of an axis other 
than the one through the Galactic poles would excise a much larger fraction of pixels 
for a rotated cluster template.}. The ensuing loss of clusters due to rotation (up to 10\% of 
all pixels fall outside the mask for a given rotation angle) introduces additional variance in 
the measured dipoles. 

Fig.~\ref{fig:rotations}  illustrates several of the artifacts introduced by PIR-13's 
rotation method. For instance, the measured monopole appears to repeat  with a 
$\sim 120^\circ$ period, and the distribution of dipoles is similarly inhomogeneous.  
If a rotated template indeed sampled random dipoles, the angular directions of these 
dipoles would have to be distributed randomly. As shown by Fig.~\ref{fig:rotations} this 
is clearly not the case; strong correlations are visible, especially for the Galactic longitude
$l$ of the dipole direction. Moreover, the amplitudes of the monopole and dipole are also 
correlated. For the T1 template, the correlation matrix of the dipole components is:
\begin{equation}
C(a_{1i},a_{1j})=\frac{\langle a_{1i} a_{1j}\rangle}{\langle a_{1i}^2\rangle^{1/2}
\langle a_{1j}^2\rangle^{1/2}}=
\left(
\begin{array}{ccc}
1 & 0.05 & -0.13 \\
0.05 & 1 & -0.26 \\
-0.13 & -0.26& 1\\
\end{array}
\right)
\label{eq:matrix}
\end{equation}
In this expression, $(i,j)=(X,Y,Z)$.
Remarkably, the dipole component along the rotation axis, Z, correlates more 
strongly with the other two components than X and Y. These correlations are larger 
than the error bars estimated from method (1), which uses the real data and random 
cluster templates, and where for 400 dipoles the off-diagonal terms never exceed 7\% 
of the diagonal terms (see eq. 10.11 and Sec. 10.3.3 in Kashlinsky, 
Atrio-Barandela \& Ebeling, 2012).

From the results shown in the top right panel of Fig.~\ref{fig:rotations} we can immediately 
compute the significance of the dipole measurement. We find the dipole measured for the 
unrotated T1 template to be significant at the 92\% confidence level; for the T2 subsample, 
the significance increases to 96\%. Large variations in the significances thus derived are 
expected, owing to the small number of dipole measurements allowed by this method. Indeed, 
PIR-13 report yet another value (89\%) when using our filter and their MCXC cluster template.

\subsection{Increment of the uncertainties in the X and Y directions.}

When the template is rotated, clusters do not move homogeneously on the 
celestial sphere. For rotations about the Z axis, the number of clusters at 
constant Galactic latitude, $b$, is fixed. Since clusters move more slowly close 
to the Galactic pole than they do near the Galactic plane, 
the Z component of the dipole is more heavily sampled than the X and Y components, resulting 
in smaller uncertainties for the former than for the latter. We can quantify this over- and 
under-sampling by using eq.~(\ref{eq:var}), which gives the error of each dipole component 
in units of the error of the monopole. 

\begin{table*}
\caption{Relative uncertainty in the measurement of the three spatial 
components of the dipole, obtained with different error-estimation methods. 
Values of unity are obtained for a uniform sampling of the sky. The error-estimation 
methods are labeled as before: (1) real data, random template; (2)  simulated data, 
cluster template; (3) real data, rotated cluster template. All methods consider only 
positions outside the mask.}
\centering
\begin{tabular}{|l|c|cc|cc|cc|}
\hline
& Method (1) & \multicolumn{2}{|c|}{Method (2) } 
& \multicolumn{2}{|c|}{Method (3) }& \multicolumn{2}{|c|}{Ratio}\\
&  & T1 & T2 & R1 & R2 & R1/T1 & R2/T2 \\
\hline
$[\langle X^2\rangle/3]^{-1/2}$ & 1.124 & 1.207 & 1.192 & 1.279 & 1.302 & 1.060& 1.092\\
$[\langle Y^2\rangle/3]^{-1/2}$ & 1.060 & 1.083 & 1.093 & 1.235 & 1.259 & 1.140& 1.152\\
$[\langle Z^2\rangle/3]^{-1/2}$ & 0.871 & 0.883 & 0.828 & 0.754 & 0.749 & 0.912& 0.906\\
\hline
\end{tabular}
\end{table*}

For clusters that are isotropically distributed on the sky the term $\langle n_i^2\rangle$ in 
eq.~(\ref{eq:var}) equals 1/3. This is the minimum variance estimator since the three 
quantities $a_{1i}$ are derived from the same data set as $a_0$, and the error  of each 
dipole component is given by $\sigma(a_{1i})=\sqrt{3}\sigma(a_0)$. 
In Table~1 we list, for all three error-estimation methods outlined before, the fractional 
increase of the errors caused by the anisotropy of both the cluster distribution and the 
sampling of the sky, relative to uniform sampling.  Note that these error estimates only 
include the effect of the geometry of the cluster template, but not the additional variances 
introduced by the limited number of dipole estimates and from the systematics intrinsic 
to the data. Hence, the figures listed in Table~1 are the minimum error that can 
be achieved for a given cluster configuration.

Note that, even for perfectly random sampling, i.e., method (1), the presence of the 
mask causes the error in the X and Y components to increase with respect to 
homogeneous sampling of the full celestial sphere, while the error in the Z component 
decreases. The same effect is evident for method (2) which results in slightly larger 
uncertainties due to the increased variance introduced by the anisotropy of the 
distribution of clusters in the sky. The by far largest deviations from uniform sampling, 
however, are induced by method (3) which employs rotations of the cluster template. 
Note (rightmost two columns in Table~1) that method (3) performs worst in the Y direction 
where it leads to a 15\% increase in the uncertainty compared to method (2). 

Fig.~\ref{fig:coadded_rotation} shows a stack of the T1 template and its 359 one-degree 
rotations, illustrating the non-uniform sampling resulting from PIR-13's rotation method. 
The observed banding, predominantly at high Galactic latitude, is a consequence of  the 
differences in both the number of clusters for different values of $b$ and in the angular 
displacement of clusters in Galactic longitude as a result of the rotation. The celestial 
sphere is sampled  inhomogeneously, increasing the errors and resulting in dipoles that 
are strongly correlated (see also eq.~(\ref{eq:matrix}), and Fig.~\ref{fig:rotations}). 

The uncertainties resulting from the three different error estimators (Table~1) can be 
translated into significances of detection for a dipole signal of given amplitude.
Fig.~\ref{fig:significance} shows these significances, based on 359 realizations 
for each of the three methods.
A dipole of amplitude $3.8\mu$K would be significant at the 97\% confidence 
level if the error estimate were obtained from 359 measurements of random dipoles 
that sample the celestial sphere perfectly uniformly.  For methods (1), (2), and (3), 
applied outside the KQ75 mask, this number is 96\%, 94\%  and less than 90\%. In 
reality, the loss of significance caused by method (3) is even higher, due to the 
correlation between the dipole components, not accounted for in these theoretical estimates.

In summary,  since it is the Y component of the dipole that dominates the Dark Flow 
signal (Kashlinsky et al.\ 2010) by increasing the error estimate on the Y-component,
method (3) (sampling of the real CMB sky by rotating the cluster template 
about the Z axis) clearly and systematically degrades the significance of a dipole 
measurement compared to methods (1) and (2). Its sampling of the celestial sphere 
is the most inefficient and most anisotropic of the three estimators, and the method 
is especially insensitive to the Y-component of the dipole. 

\section{Error estimates from CMB simulations}
\label{sec:error_sim}

Although method (3), rotation, is the most inefficient and biased error estimation 
technique discussed here, care also needs to be taken when interpreting the errors 
obtained from numerical simulations of the CMB sky, i.e., method (2). In general, 
simulated maps feature larger variances than the real sky and consequently lead to 
larger error estimates which, in turn, cause the significance of any real signal to be 
underestimated. We have identified three effects that contribute to this bias: (a) in
the filtered data the power leaks from regions outside the mask and reduces the 
variance on pixels outside the mask compared with a simulated map
(b) the power spectrum of the actual realization of the sky $C_\ell^{\rm sky}F_\ell^2$ 
is smaller than the power spectrum of the theoretical  $\Lambda$CDM concordance model 
$C_\ell^{\rm \Lambda CDM}B_\ell^2F_\ell^2$ used to generate the simulated sky, and finally, 
(c) accidental sampling of a region with non-zero dipole, if the monopole and dipole were not 
removed from the region outside the KQ75 mask, prior to computing the dipole for a random sky.

To quantify the impact of power leakage to the masked-out part of the sky we generated 
1,000 realizations of the CMB sky corresponding to the specifics of WMAP 7-year data W1 DA. 
Simulated maps were computed using Healpix $N_{side}=512$ resolution, and  multipoles 
were sampled up to $\ell_{max}=1024$. The simulated $(a_{\ell m})$ coefficients were multiplied 
by the W1 beam before transformation to real space. Next, each map was multiplied by the mask, 
converted back into spherical harmonics, filtered, and transformed once more into real space 
as described in Sec.~\ref{sec:error_theory}. For each realization, the Wiener-type filter used 
has the same functional form as specified in Sec.~\ref{sec:error_theory}, but 
$C_\ell^{\rm sky}$ is now replaced by the power spectrum of the simulated ``data'', 
$C_\ell^{\mbox{\scriptsize sim-sky}}+C_\ell^{\rm noise}$. Here 
$C_\ell^{\mbox{\scriptsize sim-sky}}$ is the power spectrum of each simulated sky, 
corrected, if necessary, for the fraction of power removed by the mask, 
and $C_\ell^{\rm noise}$ is the power spectrum of the noise. 
For each simulation we computed the rms of 
the filtered map inside and outside the mask. If in the initial data we do not
multiply by the mask, the filtered map has an average power of $\sigma=30\pm 3\mu$K.
If in the initial data we multiply by the KQ75 mask, then the power on the fraction 
of the mask outside the mask was $\sigma^{out}=28\pm 2\mu$K, while in the region excluded
by the mask it was $\sigma^{in}=15\pm 3\mu$K. This effect of power leakage is 
illustrated in Fig.~\ref{fig:maps} for one of our simulations.  We found the average 
fraction of power leaked to the mask to be $0.061\pm 0.016$. Therefore, use of 
filtered maps generated using realizations of the filtered power spectra 
not taking onto account the effect of the mask will, on 
average, increase the errors by $\sim$ 6\%.

The second effect contributing to the artificially high variance of the simulated 
CMB sky is illustrated in Fig.~\ref{fig:sigma} where we plot the mean and rms 
dispersion of the filtered power spectrum of the simulations. We define the 
cumulative variance out to a given $\ell$ as
\begin{equation}
\sigma(\ell)^2=\frac{1}{4\pi}\sum_{i=4}^{\ell}(2i+1)C_i 
\label{eq:sigmal}
\end{equation}
where $C_i$ is the filtered power spectrum. 
In Fig.~\ref{fig:sigma} the solid black line represents
the filtered power spectrum from the W1 DA data that contains both 
residual CMB and noise (solid black line), the thin red line represents
$C_\ell^{\rm \Lambda CDM,fil}=C_\ell^{\rm \Lambda CDM} B_\ell^2(F_\ell^{\rm W1})^2$, 
where $C_\ell^{\rm \Lambda CDM}$ is the theoretical power spectrum of the 
$\Lambda$CDM cosmology and $F_\ell^{W1}$ the filter constructed with 
the W1 data. Also shown is the filtered power spectrum, averaged over 1,000
simulations, (dot-dashed blue line)
and the rms dispersion around this mean (dashed blue lines). 
Let us remark that $\sigma(\ell_{max})$ is slightly 
larger than $\sigma_{\rm CMB,fil}$ of eq.~(\ref{eq:var}). We
assign all clusters a radial extent of 25\arcmin, spatial scale that 
corresponds to $\ell\sim 400$. Then, CMB residuals beyond $\ell\sim 400$
do not contribute to the variance of the monopole and dipole measured within 
cluster apertures and $\sigma_{\rm CMB,fil}\simeq \sigma(\ell\sim 300-400)$. 
As demonstrated 
in Atrio-Barandela et al.\ (2010), $\sigma_{\rm CMB,fil}$ is approximately 
$15-17\mu$K.  
Note that while in Fig.~\ref{fig:sigma} 
$C_\ell^{\rm \Lambda CDM,fil}$ is very similar to the average power spectrum 
of one thousand simulated maps (filtered using the same pipeline as applied 
to the data), the power of the actual realization of the sky $C_\ell^{sky}(F_\ell^{W1})^2$
is about 10\% smaller at $\ell=300$ (at larger $\ell$'s, it is larger due
to the noise contribution). 
Fig.~\ref{fig:sigma} shows that we live in a Universe were the residual
CMB left in the data after filtering (solid black line) is smaller
at $\ell=300-400$ than in the average sky (solid red line or dot-dashed
blue line). Since the mean of the simulations 
is larger than our actual realization of the sky 
so will be the error bars computed from simulations and by the same amount.
To properly compute the errors using simulations, care needs to be taken
to generate simulations that, on average, reproduce the residual CMB left
by the filter on the actual sky (solid black line).
We thus reach the same conclusion as 
Atrio-Barandela et al.\ (2010); adding in quadrature the power leak due to
the mask, simulations of filtered maps overestimate the errors on the monopole 
and the three dipole components by 12\% with respect to the actual data.

The third effect to consider when using simulated maps for the estimation of 
errors in the dipole measurement has already been mentioned in Sec.~\ref{sec:error_theory}. 
For each of  our thousand filtered maps,  the dipole outside the KQ75 mask must be 
removed before computing the dipole at the cluster locations, just like for the real data. 
This step is essential to ensure that  the random dipoles are computed for a 
zero-dipole sky. Keisler (2009) overlooked this crucial step, thereby artificially 
increasing his error bars. In Atrio-Barandela et al.\ (2010) we quantified that 
this omission leads to an increase in the error of the monopole close to 30\%, 
of $15-20$\% for the X and Y components of the dipole, and of 2\% for the Z 
component of the dipole.

To quantify the impact of the aforementioned biases on measurements of the 
statistical significance of the Dark Flow, we applied all three error estimation 
methods to W1 filtered data. To test the relevance of anisotropies in the cluster 
distribution, we conducted the analysis for the previously introduced T1 and T2 
templates,  but, for comparison, also for two randomly generated cluster templates 
S1 and S2 which comprise the same number of clusters as T1 and T2. We generated 
10 simulated maps of the sky and, for each of them, computed the dipole for 100 
random  positions of each template,  obtaining a total of 1,000 dipoles. Before 
computing the dipole at the cluster locations,  we removed the dipole outside the  
KQ75 mask. For simplicity, we did not include any noise or foreground residuals as 
they contribute negligibly to the final error budget. The initial power spectrum was 
$C_\ell^{\rm \Lambda CDM,fil}=C_\ell^{\rm \Lambda CDM} B_\ell^2(F_\ell^{\rm W1})^2$ 
where $C_\ell^{\rm \Lambda CDM}$  is the $\Lambda$CDM radiation power spectrum 
that best fits WMAP 7-year data, and $F_\ell^{\rm W1}$ is the Kashlinsky et al.\ (2008) 
filter for the W1 band.
To test method (3) we also computed 360 dipoles using one single map, 
by rotating the T1 and T2 templates in one-degree steps about the Z axis. We kept the 
KQ75 mask fixed during the rotation, so that pixels of the cluster templates were excised 
by the mask in the same way as the data.

The results of this exercise are shown in Fig.~\ref{fig:histograms} for the T1 and 
S1 cluster templates.  Each method assigns different statistical significance to 
the dipole measured by Kashlinsky et al.\ (2010), namely 98.6\%, 96.9\% and 
85.8\% confidence level for method (1), using the  random template S1, method 
(2), using the real cluster template T1, and method (3), using rotations of template 
T1.  Once we correct our simulations to remove the excess of power in
the simulated sky compared 
with the actual realization of the sky, as shown in Fig.~\ref{fig:sigma}, the 
significance increases to 99.6\%, 98.4\% and 90\%, respectively, showing how 
strongly the rotation method dilutes the significance. 

\section{Conclusions} 

Using Planck data and the filtering scheme of Kashlinsky et al.\ (2010), 
the PIR-13 team has detected the same dipole signal reported by 
Kashlinsky et al.\ (2010) and Kashlinsky et al.\ (2011) using WMAP data, 
but reaches a very different conclusion. Based on error estimates derived 
from simulations and rotations of the cluster template, the Planck 
Collaboration assigns the Dark Flow measurement a statistical 
significance below 90\%. 

\begin{figure*}
\centering\includegraphics[width=17cm]{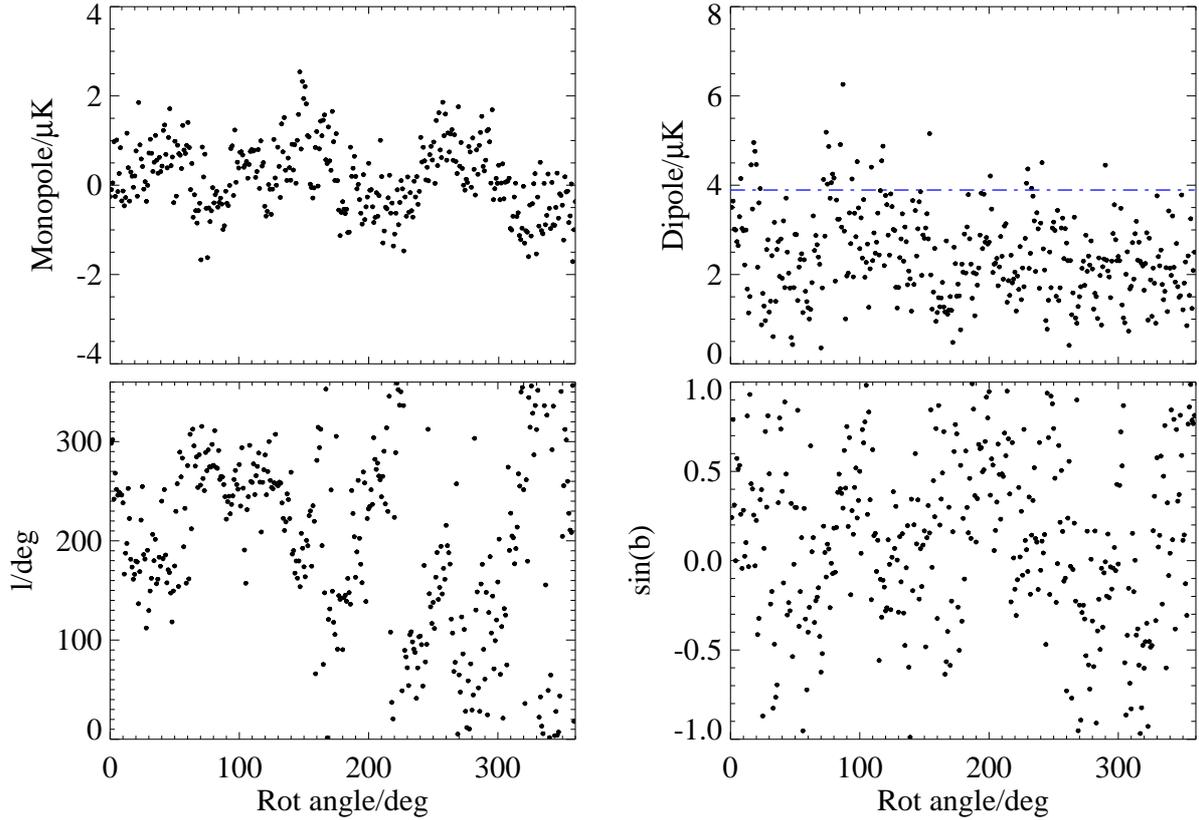}
\caption{\small Monopole and dipole components evaluated at locations obtained 
by rotating the T1 cluster template about the Z axis. The dot-dashed (blue) line 
shown in the upper right panel marks the amplitude measured at zero rotation. The 
bottom two panels show the direction of the dipole measured for a given rotation angle. 
Note the pronounced inhomogeneities in the resulting sampling of all possible random 
dipoles, as well as the 120$^\circ$ periodicity in the amplitude of the monopole.
}
\label{fig:rotations}
\end{figure*}

\begin{figure}
\centering\includegraphics[width=8.5cm]{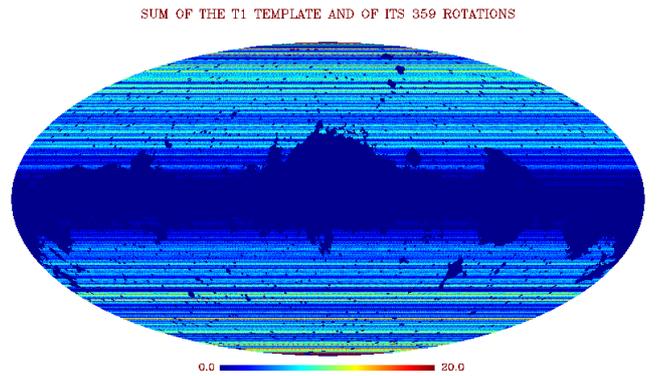}
\caption{\small Stack of the T1 cluster template and its 359 rotations about the Z axis. 
The color is proportional to the number of clusters in a given pixel 
of the template stack. All clusters were assigned an angular size of $25\arcmin$ (radius).
}
\label{fig:coadded_rotation}
\end{figure}

\begin{figure}
\centering\includegraphics[width=8.5cm]{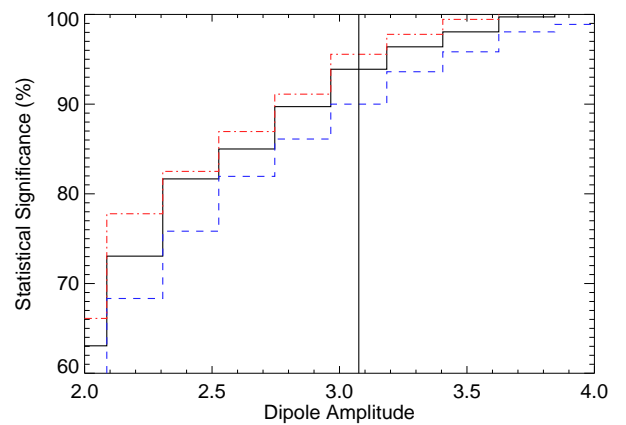}
\caption{\small 
Significance of the dipole for the three error-estimation methods. 
For a given dipole amplitude, method (1) yields the smallest errors and thus the 
highest significance (red dot-dashed line), followed by method (2) (black solid line). 
Method (3), rotations of the actual cluster template, creates the by far largest 
uncertainties and thus assigns the lowest significance.
Both method (1) and (2) were applied to 359 realizations of a random cluster 
template or a simulated CMB sky, respectively, in order to match the fixed statistics 
of 359 rotations provided by method (3).
}
\label{fig:significance}
\end{figure}

\begin{figure}
\centering\includegraphics[width=8.5cm]{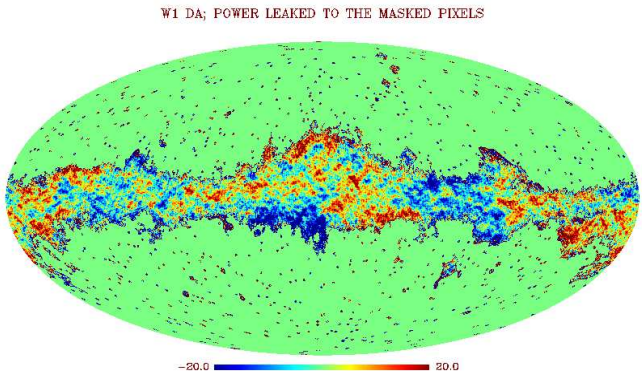}
\caption{\small Residual power in a simulated map. We use the WMAP 7-year KQ75 mask.
For clarity, the power outside the mask is set to zero and the plot is limited to the 
range $[-20,20]\mu$K. Points above/below the specified range were given the value 
of the upper / lower bound. Note the power leaked to the masked-out region of the sky.
}
\label{fig:maps}
\end{figure}

\begin{figure}
\centering\includegraphics[width=8.5cm]{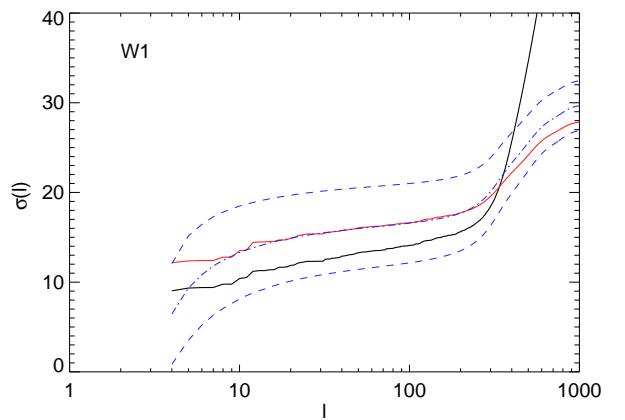}
\caption{\small 
Contribution to the rms dispersion of all multipoles up to $\ell$ (see eq~\ref{eq:sigmal})
for the WMAP W1 band.  The thick solid (black) line corresponds to WMAP data,  the thin
solid (red) line to the best-fit concordance model $C_\ell^{\rm \Lambda CDM,fil}$, 
the dot-dashed (blue) line is the mean of 1,000 simulations containing only CMB, 
and the symmetric dashed (blue) lines mark the $1\sigma$ dispersion around the mean.
}
\label{fig:sigma}
\end{figure}

\begin{figure*}
\centering\includegraphics[width=17cm]{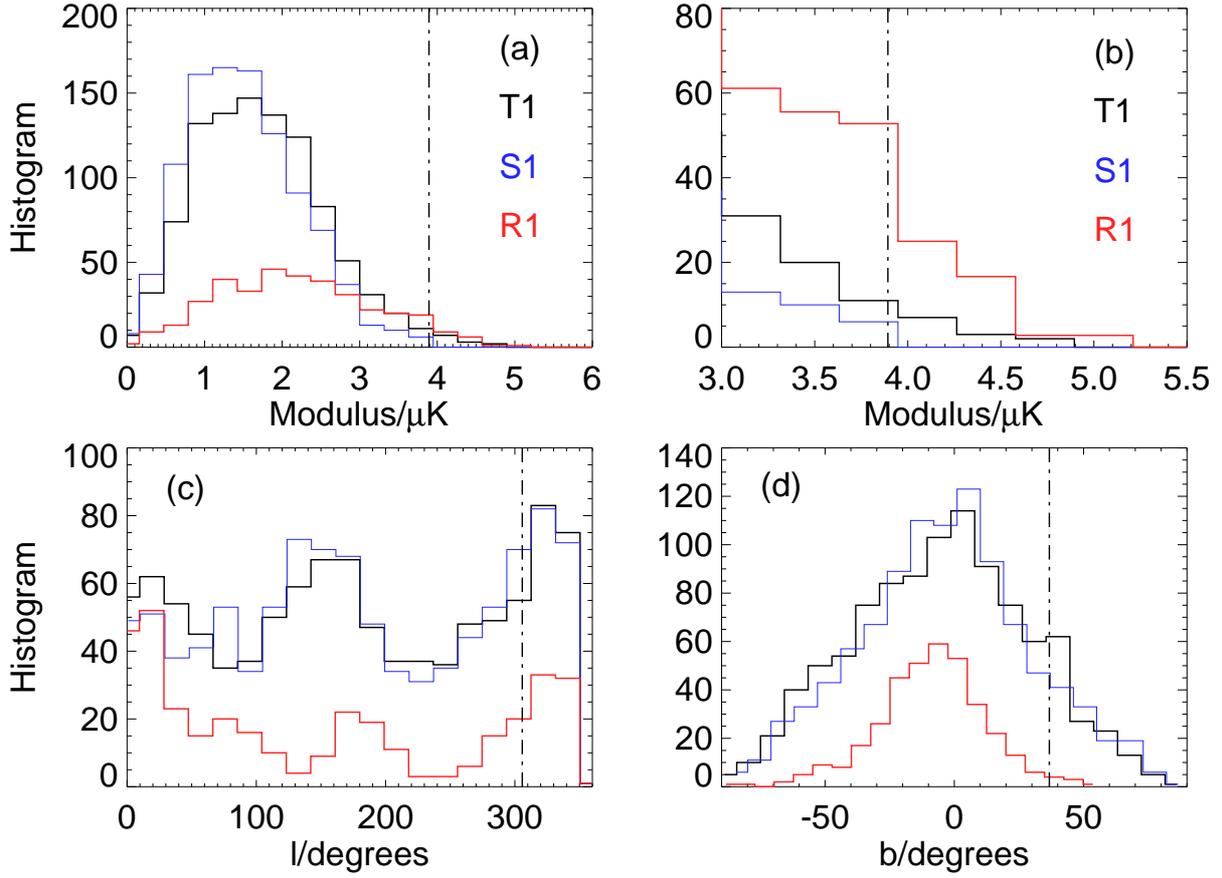}
\caption{\small 
(a) Histograms of 1,000 dipoles measured from ten simulated filtered maps for 
the W1 DA. For each map, dipoles were computed placing the template at 100 
random positions. The solid black and blue lines show the results obtained for 
the template of real clusters, T1, and for the simulated template of randomly 
positioned clusters, S1, respectively. For comparison, the solid red line shows 
the distribution of 360 dipoles computed by rotating the T1 template on
a simulated map. (b) As (a) but focusing on the $3-5.5\mu$k range and showing 
the R1 renormalized to a total of 1,000 to facilitate the comparison of the three 
error estimators. Panels (c) and (d) show the distribution of the dipole directions 
in Galactic coordinates. In all panels, the vertical dot-dashed line represents the 
measured dipole according to Kashlinsky et al.\ (2010).
}
\label{fig:histograms}
\end{figure*}

In this paper we have analyzed the error-estimation methods used in PIR-13 
and found biases that result in systematic overestimates of the uncertainty of 
the dipole measurement, and hence in a systematic underestimate of its significance. 
In particular the method of rotating the cluster template about the Z axis (devised by 
PIR-13 to account for angular correlations between clusters on the sky and all 
systematics present in the data) results in artificially inflated errors, due to several 
systematic flaws of the method: (1) since only 359  dipole measurements  are 
possible, the resulting error estimate is not statistically robust; (2) the dipole 
components are strongly correlated with each other (see Fig~\ref{fig:rotations} 
and eq.~(\ref{eq:matrix})); and (3) the sampling of the sky is highly anisotropic
(see Fig~\ref{fig:rotations}). 
Effects (2) and (3) result in errors for the X and Y components that are 5-15\% 
larger than those obtained by the error-estimation methods discussed in 
Atrio-Barandela et al.\ (2010) and Kashlinksy et al. (2012) while decreasing the 
uncertainty of the Z component of the dipole. In addition, we have shown that  
numerical simulations, as implemented by PIR-13, also overestimate the 
uncertainty of the dipole measurement. Due to cosmic variance, the actual 
realization of the sky has less power at $\ell=10-300$ than the average of an 
ensemble of simulated skies (see Fig.~\ref{fig:sigma}) which leads to errors 
derived from simulated CMB maps that are too high by about 12\%. When all 
biases are corrected for, the significance at the 90\% confidence level reported 
by PIR-13 for the Dark Flow increases to $\sim 99$\% in agreement with our 
previous estimates (Atrio-Barandela et al.\ 2010).

The strongest evidence that the measured dipole is genuinely associated with 
clusters was that it correlates with the TSZ signal measured at the cluster center 
as demonstrated in Kashlinsky et al (2010).
It is puzzling that this correlation was not observed by the 
Planck Collaboration. If the correlation is not there 
implies that either WMAP and Planck data are systematically different 
at the position of massive clusters or that the flow converges and does 
not reach as deep as the scale probed by the most massive clusters. Either 
possibility would be a very important result.

\begin{acknowledgements}
FAB acknowledges financial support from the Spanish
Ministerio de Educaci\'on y Ciencia (grants FIS2009-07238
and FIS2012-30926). Thanks are due to my collaborators
A. Edge, A. Kashlinsky, D. Kocevski and particularly to 
H. Ebeling for a careful reading of this manuscript.
\end{acknowledgements}

\pagestyle{plain}

\end{document}